\documentclass[letterpaper,10pt]{article}
\usepackage[english]{babel}
\usepackage[latin1]{inputenc}
\usepackage[pdftex]{graphicx}
\usepackage{graphicx}
\usepackage{amsfonts}
\usepackage{amssymb}
\usepackage{amsmath}
\usepackage{fancybox}

\usepackage{amsmath,amssymb}

\usepackage[pdftex,colorlinks=true,bookmarks=false,citecolor=blue,urlcolor=blue]{hyperref} %pdflatex

\begin{document}

\title{Active and passive stabilization of a high-power UV frequency-doubled diode laser}

\author{Ulrich~Eismann, Martin~Enderlein, Konstantinos Simeonidis,\\ Felix~Keller, Felix~Rohde, Dmitrijs~Opalevs, Matthias~Scholz,\\ Wilhelm~Kaenders, J\"urgen~Stuhler}

\date{}

\maketitle

\small{TOPTICA Photonics AG, Lochhamer Schlag 19, 82166 Graefelfing,  Germany}

\begin{abstract}
We present a resonantly frequency-doubled tapered amplified semiconductor laser system emitting up to 2.6\,W blue light at 400\,nm. The output power is stable on both short and long timescales with 0.12\% RMS relative intensity noise, and less than 0.15\%/h relative power loss over 16\,hours of free running continuous operation. Furthermore, the output power can be actively stabilized, and the alignment of the input beams of the tapered amplifier chip, the frequency doubling cavity and - in case of fiber output - the fiber can be optimized automatically using computer-controlled mirrors.
\end{abstract}

\section{Introduction}

High-power tunable single-frequency lasers around 400\,nm have numerous applications in research and (quantum) technology. These span from laser cooling and trapping of ions (calcium and strontium\,\cite{Eschner2003}) or atoms (erbium\,\cite{McClelland2006}, dysprosium\,\cite{Youn2010}, calcium\,\cite{Kisters1994} and chromium\,\cite{Stuhler2001}) to two-photon down conversion or even experiments with antimatter\,\cite{Eikema1999}. Furthermore, the 400-nm sources can serve as fundamental lasers for further frequency doubling to the DUV range below 200\,nm\,\cite{Scholz2013}. While Watt-class output power is required, direct diode lasers are limited to $\sim$100\,mW\,\cite{TOPTICAHP}. Frequency doubling of infrared lasers potentially delivers higher power levels, but stability becomes a major concern. Here we present a laser system consisting of a resonantly frequency-doubled  tapered amplified semiconductor laser which is capable of emitting 2.6\,Watts of single-frequency emission around 400\,nm. Based on passive and active measures, we demonstrate stable operation on timescales reaching from hundreds of nanoseconds to many days.

\section{Output power and stability}

The laser system consists of an external cavity diode laser (DL~pro), which is amplified with a semiconductor tapered amplifier chip (TA~pro) delivering up to 5\,W around 800\,nm. The almost diffraction-limited TA~pro output beam is mode matched to the second harmonic generation (SHG~pro) cavity. By choosing an input coupler reflectivity of 98.5\%, we optimize the system for SHG output power and we obtain up to 2.6\,Watts at 400\,nm with several nanometers of tunability. 
The cavity is locked to resonance by a digital control loop with a bandwidth of 30\,kHz using a slow and a fast piezoelectric transducer\,(PZT). The necessary error signal is obtained with the Pound-Drever-Hall technique\,\cite{Drever1983} from sidebands created by direct modulation of the DL~pro current at 20\,MHz. Furthermore, the digital control electronics features an automatic relock functionality. The output beam is almost diffraction limited.

\begin{figure}[htbp]
  \centering
  \includegraphics[width=\linewidth%11.5cm
  ]{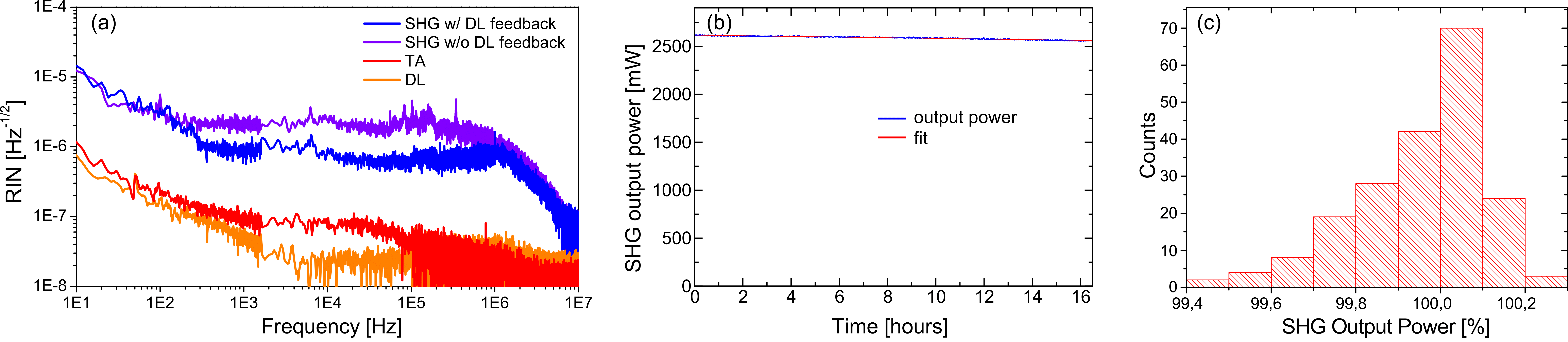}
\caption{(a) Relative intensity noise (RIN) of different laser output stages. (b) Long-term passive output stability of SHG pro light. (c) Histogram of AutoAlign results. (a)--(c): See text for details.}
\label{f_Figs}
\end{figure}

The results on a relative intensity noise (RIN) measurement of the laser system are shown in Fig.\,\ref{f_Figs}\,(a). By integrating over the full range (10\,Hz to 10\,MHz), we obtain values of 0.01\,\%\,RMS for both DL~pro and TA~pro output by accessing both beams through the respective ports. 
Due to its finite acceptance range of $f_{\rm acc}\approx 1$\,MHz, the SHG~pro cavity converts frequency noise to RIN. The DL~pro has a typical linewidth of 100\,kHz. Thus, with a 2-stage PZT lock (slow PZT for large displacements, and fast PZT), the RIN of the system at 400\,nm is increased to 0.21\,\%\,RMS (SHG w/o DL feedback in Fig.\,\ref{f_Figs}\,(a)), and leveling off at frequencies above $f_{\rm acc}$, where the cavity acts as a low-pass filter. We improve the RIN value significantly when locking the DL~pro to the SHG~pro cavity using direct feedback to the laser diode current (SHG w/ DL feedback in Fig.\,\ref{f_Figs}\,(a)), and obtain a value of 0.12\,\%\,RMS.

The SHG~pro cavity is of bow-tie type, contains a temperature controlled AR coated non-linear optical crystal and is air-sealed (typ. residual leak rate $<10^{-5}$\,mbar\,l/s). Together with the advanced mechanical design of the laser system, this yields excellent longterm stability, see Fig.\,\ref{f_Figs}\,(b). A linear fit, represented by the red line, yields a relative power drop of 0.15\%/h over more than 16 hours of continuous operation without any adjustments. Furthermore, by applying  feedback to the TA~pro current, the output power of either TA~pro stage, SHG~pro stage or the optional fiber can be actively held constant. We observe a beam pointing stability below 2\,$\mu$rad RMS over 60\,hours of continuous operation.

Even with the power stability shown in Fig.\,\ref{f_Figs}\,(b), a time-to-time realignment of the laser system may be necessary in order to achieve optimum output power. Therefore, we have developped an automatic procedure (AutoAlign) relying on a simplex algorithm and a motorization of the patented flexure mirror mounts\,\cite{Eismann2015}. The system is optimized stage-by-stage via two-mirror beam walks starting from the coupling of the DL~pro beam into the TA~pro. In a second step the TA~pro beam is aligned to the SHG~pro cavity. Finally, in case of fiber output, the coupling into fiber can be optimized.
A histogram of the results of 200 AutoAlign procedures is shown in Fig.\,\ref{f_Figs}\,(c). After deliberate misalignment, the algorithm converges to within fractions of a percent of the optimum value. The servos are switched off and completely passive when not in use for the AutoAlign procedure. Thus, no degradation of the laser parameters due to acoustic noise and heat input are observed.

\section{Conclusion and outlook}

We have presented results on a low noise, high power UV laser system allowing hands-off operation for extended periods of time. On the same platform, we have built similar systems with Watt-level power at e.g. 420\,nm, 486\,nm, 589\,nm and 671\,nm, and up to  2.7 W @ 560 nm. Frequency locking of the DL~pro master laser allows relative or absolute stabilization to laser references, and offers the potential for Hz-linewidth output\,\cite{Zhao2010} for metrology. Furthermore, these sources are efficient fundamental lasers for frequency doubling to the DUV range.

\bibliographystyle{unsrt}
\bibliography{HP_400_nm}

\end{document}